# Systematic Control of Strain-Induced Perpendicular Magnetic Anisotropy in Epitaxial Europium and Terbium Iron Garnets Thin Films

Victor H. Ortiz[1†], Mohammed Aldosary[1,2†], Junxue Li[1], Yadong Xu[1], Mark I. Lohmann[1], Pathikumar Sellappan[3], Yasuhiro Kodera[3], Javier E. Garay[3], and Jing Shi[1]

[1]Department of Physics and Astronomy, University of California, Riverside, CA 92521, USA

[2]Department of Physics and Astronomy, King Saud University, Riyadh, 11451, Saudi Arabia

[3]Department of Mechanical and Aerospace Engineering, University of California, San Diego, CA 92093, USA

We show tunable strain-induced perpendicular magnetic anisotropy (PMA) over a wide range of thicknesses in epitaxial ferrimagnetic insulator $Eu_3Fe_5O_{12}$ (EuIG) and $Tb_3Fe_5O_{12}$ (TbIG) thin films grown by pulsed-laser deposition on $Gd_3Ga_5O_{12}$ with (001) and (111) orientations, respectively. The PMA field is determined by measuring the induced anomalous Hall loops in Pt deposited on the garnet films. Due to positive magnetostriction constants, compressive in-plane strain induces a PMA field as large as 32.9 kOe for 4 nm thick EuIG and 66.7 kOe for 5 nm thick TbIG at 300 K, and relaxes extremely slowly as the garnet film thickness increases. In bilayers consisting of Pt and EuIG or Pt and TbIG, robust PMA is revealed by squared anomalous Hall hysteresis loops in Pt, the magnitude of which appears to be only related to the net magnetic moment of iron sublattices. Furthermore, the magnetostriction constant is found to be $2.7 \times 10^{-5}$ for EuIG and $1.35 \times 10^{-5}$ for TbIG, comparable with the values for bulk crystals. Our results demonstrate a general approach of tailoring magnetic anisotropy of rare earth iron garnets by utilizing modulated strain via epitaxial growth.

†These are co-first authors.
*Correspondence to: jing.shi@ucr.edu.

Ferrimagnetic insulators (FMI) have recently attracted a great deal of attention in spintronics community. On one hand, they serve as the source of pure spin currents [1] [2] [3] [4] [5], induced ferromagnetism in metals [6], graphene [7], topological insulators [8] [9], and spin Hall magnetoresistance [10] [11] [12] [13]. On the other hand, they also serve as an excellent medium for magnon spin current transport with a long decay length [14] [15]. Among various FMIs, the rare earth iron garnets (REIGs) form an interesting family. The ferrimagnetic order in REIGs results primarily from the antiferromagnetic interaction between unequal number of $Fe^{3+}$ ions on the tetrahedral and octahedral sites in a unit cell. The $RE^{3+}$ ions can have unfilled 4f-shells and therefore a finite magnetic moment that is antiferromagnetically coupled with the tetrahedral $Fe^{3+}$ moment. REIG thin films have many attractive attributes for practical applications: high Curie temperature ($T_c$ >500 K), relatively large band gaps (~2.8 eV), chemical stability, and compatibility for being incorporated into various heterostructures.

REIG thin films are often grown by pulsed laser deposition (PLD). Under strain-free conditions, the magnetization of a REIG thin film lies in the film plane because the magnetocrystalline anisotropy is generally smaller than the shape anisotropy which favors the in-plane orientation. However, due to relatively large magnetostriction constant $\lambda$, the strain-induced magnetic anisotropy energy can be even more important in thin films. This property gives epitaxial growth of REIG films a unique advantage in controlling magnetic anisotropy. Depending on the sign of $\lambda$, suitable substrates can be chosen to not only manipulate the magnitude, but also the sign of the total magnetic anisotropy energy, therefore the orientation of the magnetization vector. For example, for positive (negative) $\lambda$, compressive (tensile) strain is required to drive the magnetization normal to the film plane, which can be accomplished by controlling lattice mismatch in the pseudomorphic growth regime. The same mechanism was used for ferromagnetic metal thin films, but the interfacial strain quickly relaxes as the film thickness increases, consequently the so-called spin reorientation transition occurs only at some very small thickness (e.g., a few monolayers) [16] [17]. REIGs in general have larger Burger's vectors, **b** which give rise to larger dislocation formation energies than metals (energy scales with $|\boldsymbol{b}|^2$); therefore, the interfacial strain at garnet interfaces can extend to a larger thickness range, which makes the film thickness an additional knob to control the magnetic anisotropy in REIG films.

In order to orient the magnetization normal to the film, the perpendicular magnetic anisotropy (PMA) field $H_\perp$ must be positive and larger than the demagnetizing field. For the case of strained films grown on (100) and (111)-oriented substrates, $H_\perp$ is given by the following equations

$$H_\perp = \frac{2K_1 - 3\lambda_{100}\sigma_\parallel}{M_s}, \quad \text{and} \quad (1)$$

$$H_\perp = \frac{-4K_1 - 9\lambda_{111}\sigma_\parallel}{3M_s}, \tag{2}$$

where $K_1$ is the first-order cubic anisotropy constant, $\sigma_\parallel$ is the in-plane stress, $M_s$ the saturation magnetization and $\lambda_{lmn}$ is the magnetostriction constant for the film grown in the [$lmn$] direction. From these equations, it follows that $H_\perp$ can be controlled by tuning the in-plane stress of the film, which can be achieved by controlling growth.

In previous experiments [13] [18], it has been demonstrated that TIG (it will be referred as TmIG in this paper to avoid confusion with TbIG) can acquire strong PMA by inducing an interfacial tensile strain, since the magnetostriction constant for TmIG is negative for films grown on substituted gadolinium gallium garnet (SGGG) in [111] direction. For the cases of $Eu_3Fe_5O_{12}$ (EuIG) and $Tb_3Fe_5O_{12}$ (TbIG), the magnetostriction constants at room temperature are $\lambda_{100} = 21\times10^{-6}$, $\lambda_{111} = 1.8\times10^{-6}$ for EuIG; and $\lambda_{100} = -3.3\times10^{-6}$, $\lambda_{111} = 12\times10^{-6}$ for TbIG [19]; therefore, a compressive strain is required for all cases except TbIG (001). Given that a reasonable strain (<1%) can be accommodated for pseudomorphic growth of the film on the substrate, the candidates chosen for this study are gadolinium gallium garnets (GGG) in different orientations, i.e., GGG(001)/EuIG and GGG (111)/TbIG. The lattice mismatch between these REIG films and GGG gives rise to the compressive strain that is needed for a strong PMA field. Since the strain relaxes in thicker films, the average strain in films determines the magnetic anisotropy. Therefore, we accomplish the full anisotropy tuning by leveraging both the substrate structure and the REIG film thickness.

Thin films were grown by PLD from targets densified from powders synthesized using similar methods as described before [20]. High quality ultra-flat EuIG and TbIG films, with thickness ranging from 4 nm to 180 nm for EuIG, and 5 nm to 100 nm for TbIG, were deposited on (001)- and (111)-oriented GGG substrates respectively. After the standard cleaning process, the substrates were baked at ~ 220°C for five hours with a base pressure < $10^{-6}$ Torr before EuIG or TbIG deposition. After this annealing process, the substrates were then annealed at ~ 600°C under a 1.5 mTorr oxygen pressure with 12% (wt. %) ozone for 30 minutes; then under these oxygen and temperature conditions, a 248 nm KrF excimer laser pulse was set to strike the target with a power of 156 mJ and at a repetition rate of 1 Hz. After deposition, the films were annealed *ex situ* at 800°C for 200 seconds under a steady flow of oxygen using rapid thermal annealing (RTA).

To characterize the structural properties of the deposited garnets, reflection high energy electron diffraction (RHEED) was used to track the evolution of the film growth. Right after deposition, RHEED shows the absence of any crystalline order. After the *ex situ* RTA process, RHEED patterns appear for both EuIG and TbIG, revealing a single crystal structure for all the samples (Fig. 1 a). Atomic force microscopy (AFM) was performed on all grown samples, indicating uniform and atomically flat films with low root-mean-square (RMS) roughness (<2 Å RMS) and with no pinholes observed (Fig. 1b). The absence of three-dimensional islands on the surface from AFM measurements confirms the uniformity of the thin films.

X-ray diffraction (XRD) was performed on all the samples to further confirm their crystalline structure, using a PANalytical Empyrean diffractometer with Cu $K_\alpha$ radiation and a Ni filter, at room temperature in 0.002° steps in the $2\theta$ range of 10°-90°. For the EuIG samples, two main peaks for EuIG and GGG were observed, corresponding to the (004) and (008) Bragg peaks, while for the TbIG samples one main peak for TbIG and GGG is observed, corresponding to the (444) Bragg peak, thus confirming epitaxy and the single crystal structure of both films. For both EuIG and TbIG cases, no secondary phases were observed. These results (XRD, AFM and RHEED) combined corroborate the high quality of the obtained films.

Magnetization hysteresis curves ($M$ vs. $H$) were obtained on the grown films using a vibrating sample magnetometer (VSM) at room temperature with the applied magnetic fields normal and parallel to the plane. From the raw data, the linear paramagnetic background from the GGG substrate was subtracted. For the EuIG samples, a clear easy-axis loop can be observed for fields perpendicular to the plane for thickness $t$ up to 38 nm, while the in-plane hysteresis loop shows a hard axis behavior, thus indicating the strong PMA in these samples. For t > 56 nm, a transition from easy- to hard-axis loop can be observed for fields perpendicular to plane. This behavior can be attributed to the gradual relaxation of the strain of the films as the thickness increases, until $H_\perp$ is comparable with or less than the demagnetizing field (see Supplementary Material).

For the case of TbIG films, an easy-axis hysteresis loop can be observed in all films up to 100 nm for magnetic fields perpendicular to the plane, and a hard-axis loop for in-plane magnetic fields. In contrast to the EuIG thin films case, the TbIG thin films preserve the strong PMA over the entire thickness range (up to 100 nm). As the product of the magnetostriction constant and in-plane lattice strain ($\lambda_{lmn}\sigma_\parallel$) is generally larger in EuIG than in TbIG, the apparently larger $H_\perp$ in TbIG can be attributed to the smaller saturation magnetization in TbIG compared to EuIG (to be discussed below).

The hysteresis loops for 38 nm EuIG and 100 nm TbIG films are included in the Supplementary Material. These loops clearly indicate PMA in the films. For thinner films, i.e., $t < 38$ nm for EuIG and $t < 100$ nm for TbIG, both REIGs have stronger PMA. The saturation magnetization for all films is summarized in Table I. For EuIG films with $t < 14$ nm and for TbIG films with $t < 20$ nm, the magnetic moment signal is too small to be resolved by the VSM due to the large background signal from GGG and their magnetization data are not included. The average saturation magnetization for EuIG is $4\pi M_s = (913 \pm 7)$ G, which is 23.5% smaller than the reported value for bulk EuIG ($4\pi M_s = 1192.83$ G) [22], which might be caused by a variation in stoichiometry as it has been observed in similar studies [21]. For the case of TbIG, the average saturation magnetization is $4\pi M_s = (234 \pm 5)$ G, which is only 3.4% below the reported value for bulk TbIG ($4\pi M_s = 242.21$ G).

For films with $t > 10$ nm, the XRD data shows that the Bragg peak corresponding to both REIGs is shifted to the left from the expected peak positions for the respective bulk crystals (Fig. 2a-b), thus indicating an elongation on the lattice parameter perpendicular to the surface, leading to a compressive in-plane strain in the lattice. Moreover, a systematic shift to higher $2\theta$ values of the diffraction peaks for both EuIG and TbIG as the thickness of the thin film increases is a direct measurement of the relaxation of the lattice parameters towards the bulk values (Fig. 2c-d), in contrast to the results obtained by Rosenberg *et al.* [22]. The surprisingly slow relaxation behavior in REIG thin films contrasts sharply with ferromagnetic metal films, demonstrating a unique magnetic anisotropy control possibility by film thickness.

In order to determine the in-plane stress $\sigma_\parallel$ in the garnet films, it is necessary to consider the elastic deformation tensor of the material. When a film with cubic crystalline structure is grown on a single crystal substrate, and assuming the material originally is isotropic, then two strain components can be considered: an in-plane biaxial strain $\epsilon_\parallel$ and an out-plane uniaxial strain $\epsilon_\perp$. These parameters are related through the elastic stiffness constants by [23] [24]

$$\epsilon_\parallel = -\frac{c_{11}}{2\, c_{12}}\, \epsilon_\perp \qquad \text{for film grown on (001) substrate,} \qquad (3)$$

$$\epsilon_\parallel = -\frac{c_{11}+2\, c_{12}+4\, c_{44}}{2c_{11}+4\, c_{12}-4\, c_{44}}\, \epsilon_\perp \qquad \text{for film grown on (111) substrate,} \qquad (4)$$

and

$$\epsilon_\perp = \frac{c - a_0}{a_0}, \qquad (5)$$

$c$ being the out-of-plane lattice parameter for the strained film, and $a_0$ the lattice parameter for the relaxed (bulk) material; $c$ can be obtained from the XRD data according to the equation $c = d_{hkl}\sqrt{h^2 + k^2 + l^2}$. For the case $t < 10$ nm, $\epsilon_\parallel$ can be obtained directly from the RHEED pattern. $\sigma_\parallel$ can be calculated then by

$$\sigma_\| = -\frac{c_{11}}{2c_{12}}\left(c_{11} + c_{12} - \frac{2c_{12}^2}{c_{11}}\right)\epsilon_\perp \qquad \text{for the (001) case, and} \qquad (6)$$

$$\sigma_\| = -6c_{44}\frac{c_{11}+2c_{12}}{2c_{11}+4\,c_{12}-4\,c_{44}}\epsilon_\perp \qquad \text{for the (111) case,} \qquad (7)$$

where $c_{ii}$ corresponds to the elastic stiffness constants. For EuIG ($c_{11}$ = 25.10×10$^{11}$ dyne/cm$^2$, $c_{12}$ = 10.70×10$^{11}$ dyne/cm$^2$, $c_{44}$ = 7.62×10$^{11}$ dyne/cm$^2$) [25] and TbIG ($c_{11}$ = 26.53×10$^{11}$ dyne/cm$^2$, $c_{12}$ = 11.07×10$^{11}$ dyne/cm$^2$, $c_{44}$ = 7.15×10$^{11}$ dyne/cm$^2$) [26], the values for $\sigma_\|$ can be calculated then from equations (6) and (7), which are listed in Table I.

An important factor is the combination of the lattice-mismatch-induced in-plane compressive strain and positive magnetostriction coefficients which can drive the magnetization perpendicular to the film plane in both EuIG and TbIG. As mentioned before, in TmIG, due to negative magnetostriction constant, tensile strain is needed to obtain PMA, which was achieved by growing it on SGGG or NGG substrates. The strong PMA in those garnet films are characterized by squared magnetic hysteresis loops for out-of-plane fields but a hard-axis behavior for in-plane fields. To quantify $H_\perp$ in EuIG and TbIG films, magneto-transport measurements in REIG/Pt bilayers are performed at room temperature. Since these REIG films are magnetic insulators, the Hall response of Pt imprints the magnetic anisotropy of the REIG films via the magnetic proximity effect and/or the spin current effect [13]. Therefore, a 5 nm thick Pt layer was sputtered into the Hall bar geometry with a length of $l$ = 600 μm and a width of $w$ = 100 μm using standard photolithography. The inset of Fig. 3(a) shows the optical image of the Hall bar shape and dimensions. Measured Hall response contains two parts: the ordinary Hall effect (OHE) which is linear in field, and the anomalous Hall effect (AHE) which is proportional to the out-of-plane magnetization. Fig. 3(a) shows sharp squared out-of-plane AHE hysteresis loops after subtraction of linear background of OHE. These loops resemble $M$ vs. $H$ loops with the easy axis character of TbIG, EuIG and TmIG taken with an out-of-plane magnetic field. To compare the AHE magnitude, we kept the REIG and Pt thicknesses constant at 30 nm and 5 nm respectively for the three bilayers. Measured AHE resistivity magnitude ($\rho_{AHE}$) for the REIG/Pt systems are 0.429, 0.529, and 0.89 $n\Omega \cdot cm$ for EuIG/Pt, TmIG/Pt and TbIG/Pt, respectively. Fig. 3(b) shows the comparison of both measured $4\pi M_s$ and $\rho_{AHE}$ from Fig. 3(a) among all three bilayers. The $4\pi M_s$ value is 1400 G for TmIG, 915 G for EuIG, and 225 G for TbIG. The difference in the magnetization is due to the different magnetic moment of the rare-earth elements which are coupled antiferromagnetically with the net $Fe^{3+}$ moment and causes partial compensation of the net $Fe^{3+}$ moment. Note that the ($4\pi M_s$) decreases from TmIG to TbIG by a factor of five whereas $\rho_{AHE}$ stays in nearly the same range for all of them. This sharp contrast clearly indicates that the $\rho_{AHE}$ of the Pt layer is correlated with the net magnetic moment of sublattices of $Fe^{3+}$ ions, which is a constant, rather than the total magnetic moment of REIGs (including $Fe^{3+}$ and $RE^{3+}$). This can be explained by the fact that the conduction electrons of Pt are primarily hybridized with the $3d$ $Fe^{3+}$ electrons which are more spatially extended than the $4f$ electrons of the rare earth elements.

Magnetic anisotropy energy of thin films in general consists of three terms and can be written as

$$K_u = 2\pi M_s^2 + K_c + K_\sigma, \qquad (8)$$

where $K_c$ is magnetocrystalline anisotropy which can be approximated by the first-order cubic anisotropy constant ($K_1$). In EuIG and TbIG, ($K_1$) is negative and ($\approx -10^4 erg/cm^3$); the term $2\pi M_s^2$ corresponds to the shape anisotropy ($\approx 10^4 erg/cm^3$) which gives the in-plane demagnetizing field; $K_\sigma$ is the strain-induced anisotropy which is determined by magnetostriction coefficient ($\lambda_{lmn}$) and in-plane strain ($\varepsilon_\parallel$). For EuIG and TbIG, $K_\sigma$ is positive and large ($\approx 10^5 erg/cm^3$) as can be seen in Table I. As a result, comparing these three anisotropies, $K_\sigma$ is at least one order of magnitude larger than $K_c$ and $2\pi M_s^2$. Therefore, to evaluate how the compressive strain influences the magnetic anisotropy, we first determine the anisotropy field from the hard-axis AHE loops and quantitatively study $t$-dependence in EuIG/Pt(5 nm) and TbIG/Pt(5 nm). The EuIG films with $t$ up to 56 nm show out-of-plane magnetization whereas those above 56 nm show in-plane magnetization, indicating that the PMA field is overcome by the demagnetizing field above this thickness. On the other hand, TbIG shows perpendicular magnetization for all films with the thickness up to 100 nm. This is primarily due to a smaller saturation magnetization value in TbIG which causes the PMA field to be dominant over the demagnetizing field in the entire thickness range. Detailed transport measurements for the Hall resistivity ($\rho_{AHE}$) and longitudinal magnetoresistance (MR) ($R_{xx}$) responses are performed with *in-plane field sweeps* to reach in-plane magnetic field saturation value ($H_S^{ip}$) in which both $\rho_{AHE}$ and $R_{xx}$ signals saturate. Fig. 3(c & d) show two examples of the $H_S^{ip}$ extraction procedure on EuIG (20 nm)/Pt (5 nm) and TbIG (60 nm)/Pt (5 nm) with perpendicular magnetization easy axis from the Hall and MR which have the same values. Dashed lines in Fig. 3 (c & d) define the in-plane $H_S^{ip}$ for 20 nm thick EuIG with saturation of 14.50 kOe and for 60 nm thick TbIG with saturation of 17.50 kOe. From transport measurements, $H_\perp$ is extracted from the following relation

$$H_S^{ip} = H_\perp - 4\pi M_s \qquad (9)$$

for EuIG $t < 56$ nm and for all TbIG thicknesses. However, EuIG films need special care to determine the $H_\perp$ for $t > 56$ nm. Those films have in-plane magnetic anisotropy, and the Hall voltage may contain a planar Hall component when an in-plane field is applied. In this case, $H_\perp$ is negative, and an out-of-plane saturation field

$H_S^{op}$ is measured instead (see Supplementary Material). Then equation (9) must be replaced by

$$H_S^{op} = 4\pi M_s - H_\perp. \qquad (10)$$

Figures 4a and 4b show the plot $H_\perp$ as a function of film thickness $t$; $H_\perp$ follows a $1/(t+t_o)$ behavior (fitted as $H_\perp = \left(\frac{691.05}{t+13.69} - 8.48\right) kOe$ for EuIG, $H_\perp = \left(\frac{4472.76}{t+48.50} - 24.36\right) kOe$ for TbIG), as it has been observed in other works [27]. As described in equations (1) and (2), $H_\perp$ has a linear relation with the in-plane strain $\epsilon_\parallel$, and is expected to follow the same behavior as the out-of-plane strain, $\varepsilon_\perp$, according to equations (3) and (4). There are two regimes for the effect of lattice mismatch in thin films [28]: for film thickness $t$ below certain critical value $t_c$, the film will be pseudomorphic and the strain is given by $\eta = \frac{a_f - a_s}{a_s}$, while for $t > t_c$, the strain relaxes as $\varepsilon_\parallel \propto \eta \frac{t_c}{t}$. The insets in figures 4a and 4b show that for the measured thickness ($t > 4$ nm), the $1/(t+t_o)$ behavior is observed, indicating that the strain in the lattice relaxes over the entire thickness range. The fitted equation for the out-of-plane strain versus thickness is consistent with this $1/(t+t_o)$ behavior, with a residual value of 0.34% for EuIG and 0.01% for TbIG as $t \to \infty$. The relatively large residual strain in the thick EuIG film limit suggests that the PLD films differ somewhat from the bulk crystals. The difference may be caused by oxygen deficiency or other defects in PLD films, which may account for the relatively low saturation magnetization as mentioned earlier.

Figures 4c and 4d show $H_\perp$ as a function of $\epsilon_\parallel$ for both EuIG and TbIG respectively, where a linear relation was observed. A least square fitting was performed and the magnetostriction coefficient $\lambda$ was calculated. For EuIG, it was found that $\lambda_{100EuIG} = (2.7 \pm 0.1) \times 10^{-5}$, which is about 29% larger than the reported value ($\lambda_{100} = 2.1 \times 10^{-5}$), while for TbIG, it was found that $\lambda_{111TbIG} = (1.35 \pm 0.06) \times 10^{-5}$, being only 12% larger than the literature value ($\lambda_{111} = 1.2 \times 10^{-5}$) [19], these variations may be attributed to the difference in growth conditions of both bulk and thin film samples which result in slightly different material properties and additionally the difference in measurement techniques.

In summary, we have utilized compressive in-plane strain to control the perpendicular magnetic anisotropy in coherently strained epitaxial EuIG (001) and TbIG (111) thin films. Electrical measurements performed on Pt Hall bars fabricated on these films show squared AHE hysteresis loops which are primarily sensitive to the net magnetic moments of $Fe^{3+}$. The PMA field relaxes extremely slowly as the ferrimagnetic insulator thickness increases. Our experimental results demonstrate a full control of magnetic anisotropy of REIG ferrimagnetic insulators using epitaxial growth.

SUPPLEMENTARY MATERIAL

See Supplementary Material for magnetic hysteresis loops of selected EuIG and TbIG films measured by VSM and transport measurements for saturation field determination.

ACHKNOWLEDGEMENTS

    We would like to thank Yawen Liu and Chi Tang for many fruitful discussions, and Dong Yan and John Butler for their technical assistance. The work was supported as part of the SHINES, an Energy Frontier Research Center funded by the US Department of Energy, Office of Science, Basic Energy Sciences under Award No. SC0012670.

**Figures**

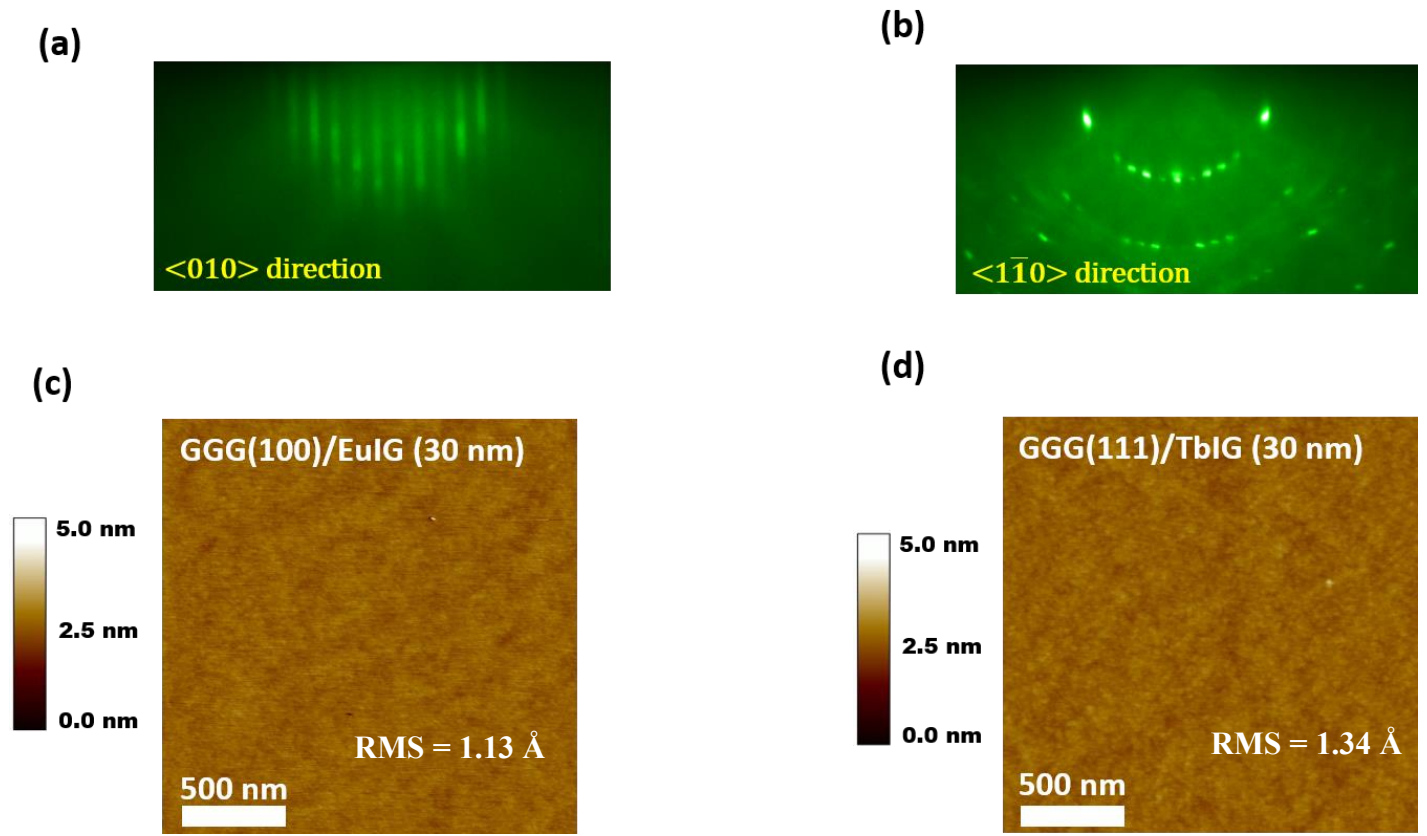

Figure 1. (a) and (b): RHEED patterns of (a) EuIG along <010> direction and (b) TbIG along <1$\bar{1}$0> direction. Both show single crystal structure after annealing. (c) and (d): $2\mu m \times 2\mu m$ AFM surface morphology scans of EuIG (30nm) thin film (c) with RMS roughness of 1.13 Å and TbIG (30 nm) thin film (d) with RMS roughness of 1.34 Å.

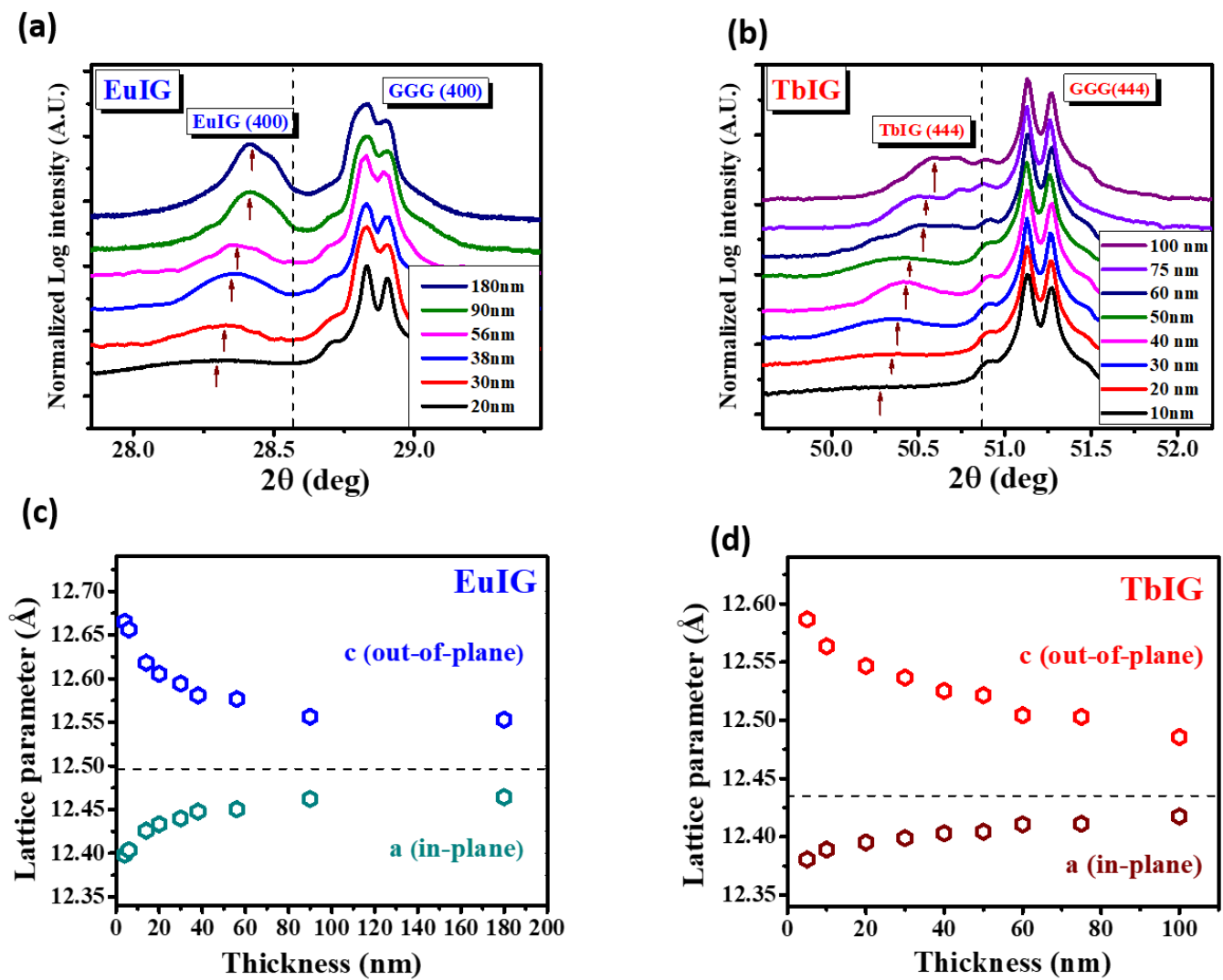

Figure 2. (a) and (b): Normalized semi-log plot of $\theta$-$2\theta$ XRD scans of (a) EuIG films of thickness t = 20, 30, 38, 56, 89, 178 nm grown on GGG (100) and (b) TbIG of thickness t= 10, 20, 30, 40, 50, 60, 75 and 100 nm grown on GGG(111) substrate. The dashed line shows the 2θ positions of the bulk material. The arrows indicate the position evolution of (004) peak for EuIG and (444) peak for TbIG as strain relaxes. (c) and (d): Thickness dependence of the out-of-plane lattice constant $c$ and in-plane lattice constant $a$ for EuIG films on GGG(001) (c) and TbIG on GGG(111) (d). The dashed line represents the bulk lattice constant (a=12.497 Å) for EuIG and (a= 12.435 Å) for TbIG.

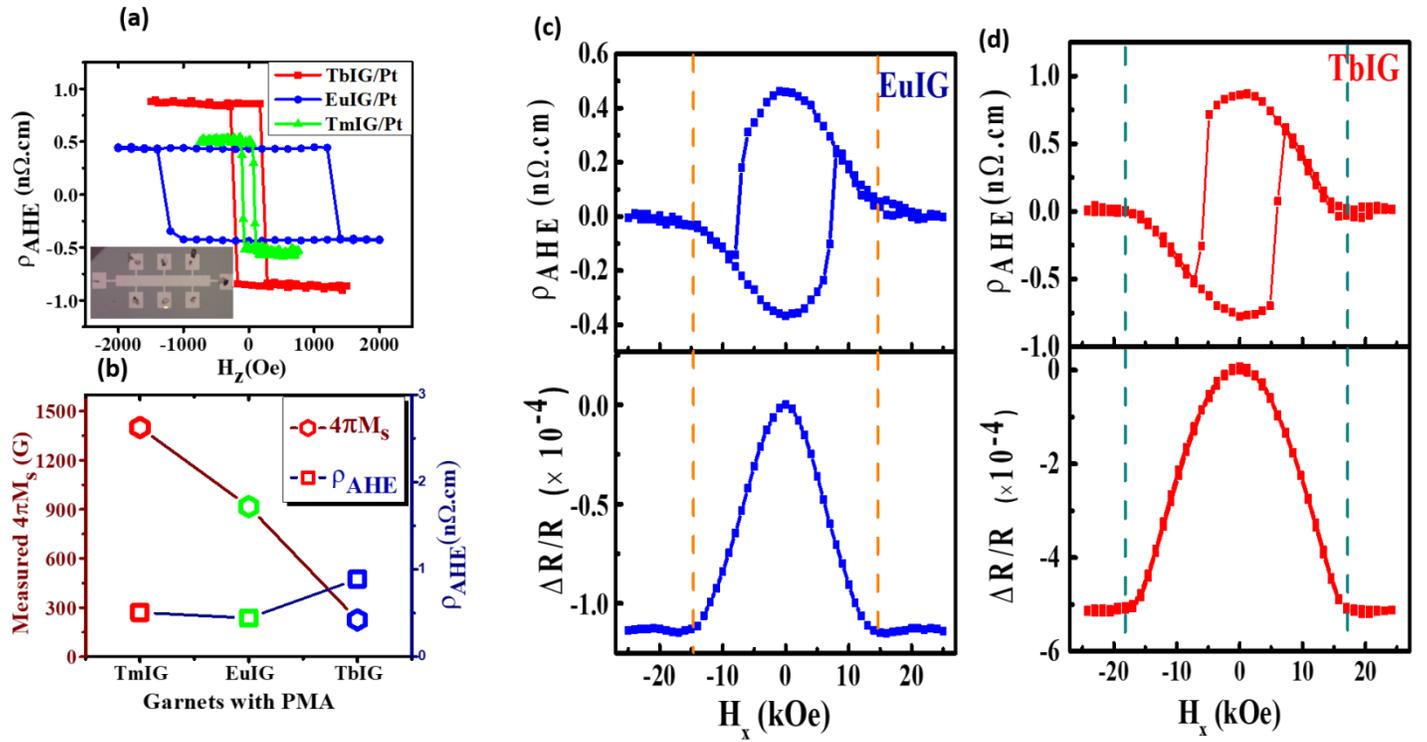

Figure 3. (a) The anomalous Hall resistivity measured with an out-of-plane field sweep for three PMA garnets (TbIG (30 nm)/Pt(5 nm), EuIG(30 nm)/Pt(5 nm) and TmIG(30 nm)/Pt(5 nm)). Inset: Optical microscope image of Hall-bar device with a length of L= 600 μm and a width of w= 100 μm. (b) Measured magnetization values ($4\pi M_s$) and anomalous Hall resistivity magnitude as a function of PMA garnets (TbIG, EuIG and TmIG). (c) and (d) Hall resistivity ($\rho_{AHE}$) and longitudinal magnetoresistance ratio ($\Delta R/R$) as a function of in-plane magnetic field for (c) EuIG (20 nm)/Pt(5 nm) and (d) TbIG (60 nm)/Pt(5 nm). The dashed lines represent the saturation magnetic field for in-plane geometry.

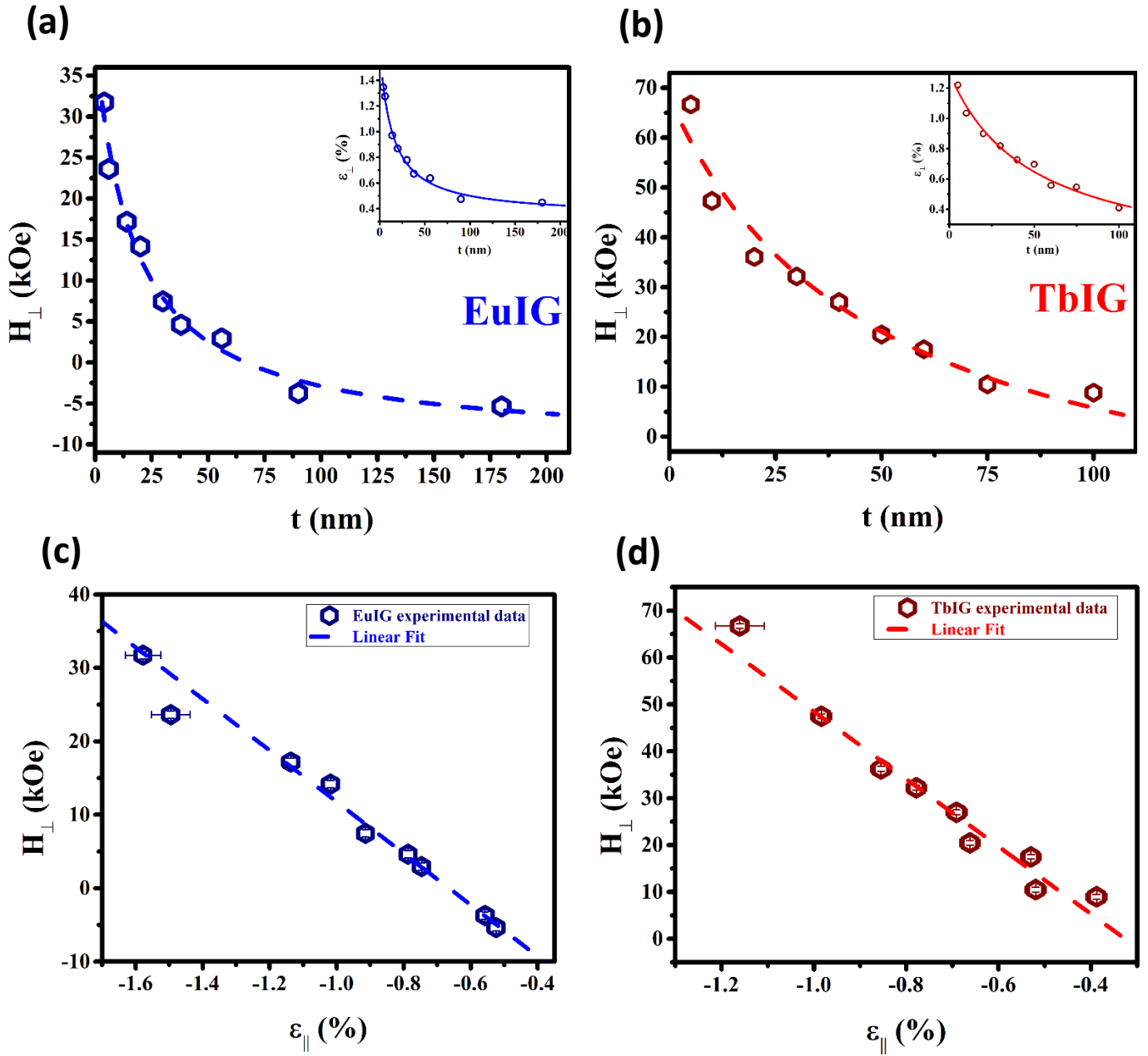

Figure 4. (a) and (b): $H_⊥$ vs. $t$ for (a) EuIG and (b) TbIG. In the inset the out-of-plane strain $ε_⊥$ as a function of film thickness, showing the corresponding $1/t+t_o$ behavior for the measured interval. (c) and (d): $H_⊥$ as a function in-plane strain $ε_∥$ for (a) EuIG and (b) TbIG. The dashed line is linear fit to obtain thin film magnetostriction constant ($λ_{100}$) for EuIG and ($λ_{111}$) for TbIG.

| GGG (001)/EuIG | | | | | | | | |
|---|---|---|---|---|---|---|---|---|
| Thickness (nm) | c (Å) | $\epsilon_\perp$(%) | $\epsilon_\parallel$(%) | $\sigma_\parallel \times 10^{10}$ dyne/cm$^2$ | $4\pi M_s$ (Gauss) | $H_{sat}$ (kOe) | $H_\perp$ (kOe) | $K_u \times 10^5$ (erg/cm$^3$) |
| 4 | 12.665 | 1.35 | -1.58 | -4.21 | -- | 32.00 | 32.91* | 11.96* |
| 6 | 12.656 | 1.28 | -1.50 | -3.99 | -- | 24.00 | 24.91* | 8.719* |
| 14 | 12.618 | 0.97 | -1.14 | -3.03 | 916 | 17.50 | 18.42 | 6.375 |
| 20 | 12.605 | 0.87 | -1.02 | -2.72 | 899 | 14.50 | 15.40 | 5.185 |
| 30 | 12.594 | 0.78 | -0.91 | -2.44 | 932 | 7.75 | 8.68 | 2.874 |
| 38 | 12.581 | 0.67 | -0.79 | -2.10 | 911 | 5.00 | 5.91 | 1.811 |
| 56 | 12.577 | 0.64 | -0.75 | -1.99 | 879 | 3.25 | 4.13 | 1.137 |
| 90 | 12.556 | 0.48 | -0.56 | -1.49 | 927 | 3.50 | -2.57 | -1.290 |
| 180 | 12.553 | 0.45 | -0.52 | -1.40 | 927 | 6.00 | -5.07 | -1.870 |
| GGG (111)/TbIG | | | | | | | | |
| Thickness (nm) | c (Å) | $\epsilon_\perp$(%) | $\epsilon_\parallel$(%) | $\sigma_\parallel \times 10^{10}$ (dyne/cm$^2$) | $4\pi M_s$ (Gauss) | $H_{sat}$ (kOe) | $H_\perp$ (kOe) | $K_u \times 10^5$ (erg/cm$^3$) |
| 5 | 12.587 | 1.22 | -1.16 | -3.71 | -- | 66.50 | 66.73* | 6.213* |
| 10 | 12.564 | 1.03 | -0.98 | -3.14 | -- | 47.25 | 47.48* | 4.421* |
| 20 | 12.547 | 0.90 | -0.86 | -2.73 | 253 | 36.00 | 36.25 | 3.652 |
| 30 | 12.537 | 0.82 | -0.78 | -2.48 | 223 | 32.00 | 32.22 | 2.852 |
| 40 | 12.525 | 0.73 | -0.69 | -2.21 | 242 | 26.75 | 26.99 | 2.598 |
| 50 | 12.522 | 0.70 | -0.66 | -2.11 | 221 | 20.25 | 20.47 | 1.803 |
| 60 | 12.504 | 0.56 | -0.53 | -1.69 | 230 | 17.25 | 17.48 | 1.603 |
| 75 | 12.503 | 0.55 | -0.52 | -1.66 | 247 | 10.25 | 10.50 | 1.030 |
| 100 | 12.486 | 0.41 | -0.39 | -1.24 | 222 | 8.75 | 8.97 | 0.792 |

Table I. Structural and magnetic property parameters for epitaxially strained EuIG films with thickness $4 \leq t \leq 180$ nm on GGG (001) and TbIG films with thickness $5 \leq t \leq 100$ nm on GGG (111). The $H_\perp$ and $K_u$ values marked with (*) were calculated using the average $4\pi M_s$ values.